\documentclass[12pt]{iopart}
\pdfoutput=1
\usepackage{graphicx}
\usepackage{color}
\usepackage{cite}   
\DeclareGraphicsExtensions{.pdf,.png,.jpg}

\newcommand{\keywords}[1]{\begin{indented}
   \item[Keywords: ]\rm\raggedright #1
   \end{indented}}
\begin{document}
\title{Low-coverage heteroepitaxial growth with interfacial mixing}
\author{K.~I.~Mazzitello, L.~M.~Delgado}
\address{Facultad de Ingenier\'{\i}a,
             Universidad Nacional de Mar del Plata, \\
Av.~J.~B.~Justo 4302, 7600 Mar del Plata,
         Argentina}

\author{J.~L.~Iguain}
\address{Instituto de Investigaciones F\'{\i}sicas de Mar del Plata (IFIMAR) 
and\\
Departamento de F\'{\i}sica FCEyN,
Universidad Nacional de Mar del Plata,\\
De\'an Funes 3350, 7600 Mar del Plata, Argentina}

\begin{abstract}
We investigate the influence of intermixing on heteroepitaxial
growth dynamics, using a two-dimensional point island model, expected to
be a good approximation in the early stages of epitaxy.
In this model, which we explore both analytically and numerically,
every deposited B atom diffuses on the surface with  diffusion constant
$D_{\rm B}$, and can exchange with any A atom of the substrate at
constant rate. There is no exchange back, and emerging atoms diffuse
on the surface with diffusion constant $D_{\rm A}$.   When any two
diffusing atoms meet, they nucleate a point island. The islands neither
diffuse nor break, and grow by capturing other diffusing atoms. 
The model leads to an island density governed by the diffusion of 
one of the species at low temperature, and by the diffusion of the other 
at high temperature. 
We show that these limit behaviors,
as well as intermediate ones, all belong to the same universality
class, described by a scaling law.
  We also show that the island-size distribution is 
self-similarly described 
by a dynamic scaling law in the limits where only one diffusion
constant is relevant to the dynamics, and that this law is
affected when both $D_{\rm A}$ and $D_{\rm B}$ play a role.
\end{abstract}


 \maketitle
%%%%%%%%%%%%%%%%%%%%%%%%%%%%%%%%%%%%%%%%%%%%%%%%%%%%%%%%%%%%%%%%%%%%%%%%%%%%
\section{Introduction}
\label{intro} 
Much of recent research on heteroepitaxial growth is focused to developing nanometer-scale devices 
with novel properties. 
Quality, performance and lifetime of these
devices are determined by the purity, structural perfection and homogeneity of
the epitaxial layers. Surface flatness and interface abruptness obtained through epitaxial crystal growth depend on the relative values of the interfacial energy and the surface free energy of the substrate and the film, under equilibrium conditions. However, in most cases thin films are grown far away from thermodynamic equilibrium, leading to kinetically controlled processes. 
  Surface structures, thus depend in a very complicated way on 
several variables, which in simplest models include the deposition flux,  
the mobility the deposited particles,  nucleation and detachment rates, and 
the interfacial energy between substrate and epitaxial film
 \cite{mich04,evans_review}.
Besides these processes, an additional mechanism, shown to be important in
 many cases of heteroepitaxial growths, is that of exchange, in which a 
deposited atom becomes embedded into the substrate and a substrate atom is 
removed. Exchange leads to growth of islands of mixed composition. 
Intermixing is specially undesired in case of magnetic materials, as it 
produces a decrease in the interface magnetization with respect to expected. 
It has been reported that 
V~\cite{cite1Kang04}, Fe~\cite{cite2Kang04}, Co~\cite{cite3Kang04}, 
Ni~\cite{cite4Kang04}, Cr~\cite{Kang04}, Ir~\cite{Niehus99} intermix with Cu 
atoms at their interfaces forming alloy layers, and for instance, the 
average magnetic moment of 4 mono-layer Ni film on Cu(001) is half of that in 
the bulk Ni, as detected by X-ray magnetic circular dichroism 
measurements~\cite{cite6Kang04}.\par

It is not at all surprising that exchange occurs for two elements that are 
completely miscible like, for instance, Au and Ag \cite{Chambliss95}. 
Deposition of Au on Ag(110) forms alloy-like structures that are not 
energetically costly and the comparatively open atomic geometry of an fcc(110) 
surface makes place exchange 
possible with fairly small bond distortions. However, intermixing of the 
constituents may also well occur 
for bulk immiscible systems. The phase diagram of the Ir-Cu system shows a 
massive miscibility gap. At 
temperatures up to around 1000 K only 3 at.~\% Ir appears soluble in  Cu and 
in the reverse case only 1 at.~\% Cu in Ir~\cite{7Niehus99}. No intermixing 
would be expected for these elements, at least at 
low temperatures. As the surface free energy of Ir is considerably higher 
than of Cu (3 $J/m^2$ and 1.83 $J/m^2$, respectively \cite{8Niehus99}), when 
Ir is deposited on Cu one should observe 3D 
growing clusters composed only of Ir atoms.  However, experimental results for 
Ir on Cu(100) unequivocally show intermixing, even at room 
temperature~\cite{Niehus99}.
Thus, structures resulting from heteroepitaxy are often complex and difficult 
to predict from bulk material parameters.\par

A common fact of heteroepitaxial systems with intermixing is that the surface 
free energy of the deposited atoms is higher than the substrate 
one~\cite{15Niehus99,16Niehus99,17Niehus99,18Niehus99,19Niehus99,20Niehus99}. 
At high enough coverage, this results in step roughening; which might then be 
considered as an indicator of intermixing. However, well before 3D islands arise
 on the substrate, the question remains about what are the effects of
intermixing at the early stages of heteroepitaxy. 
  The case where deposited
monomers can react with the substrate was studied in \cite{Chambliss94,Zang95}.
After an irrevesible exchange, these atoms become immobile and act as centers
of nucleation, which form inclusions in the substrate. The emerging atoms 
become mobile but are assumed to adhere to a step elsewhere, and play no
role in the dynamics.
The effects of intermixing on the structures formed on the substrate where
analyzed in~\cite{Bier04}. In this work, islands nucleate by the encounter of any pair of diffusing atoms, and the authors studied the properties
of the concentration and the spatial correlation of substrate atoms which 
become part of the islands. However, in order to keep
the analysis simple, they assume that both species diffuse equally fast on the surface.

In this paper, we address the problem of growth dynamics with intermixing
according to a model of point islands ( 
which occupy a single site\cite{6PRE,7PRE}), expected to be a valid approximation at 
low enough coverages. The exchange rate is entered as a parameter,   and both constants
of diffusion are taken into account
Our approach is two-fold: theoretical analysis and numerical 
Monte Carlo (MC) simulations.

The paper is organized as follows.  In section \ref{model} we define the model. 
We consider two species of atoms, and the dynamics depend on the intermixing 
and deposition rates as well as on the diffusion 
constants of both species. The main results are presented in section 
\ref{discussion}. In \ref{Island density}, we analyze the behavior of the 
density of islands. The composition of interface at low coverage is studied 
in \ref{Surface composition}, and the results of simulations are compared with 
experiments. In \ref{Mean-field}, we state and solve mean-field 
evolution equations for island an monomer densities.
These equations lead to a scaling form for the island density,  
described in \ref{Scaling}.   A reduced form of the dynamic scaling 
of the island-size distribution is presented in \ref{Dynamic Scaling}. 
Finally, in section \ref{Conclusions}, we state our conclusions.

%%%%%%%%%%%%%%%%%%%%%%%%%%%%%%%%%%%%%%%%%%%%%%%%%%%%%%%%%%%%%%%%%%%%%%%%%%%%

\section{The model}
\label{model}

\begin{figure}[!ht]
\begin{center}
\includegraphics[width= \linewidth]{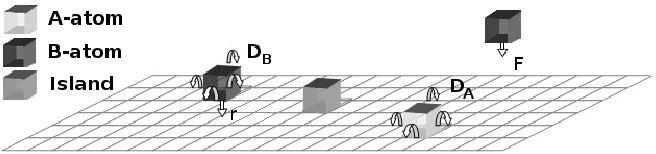}
\end{center}
\caption{Schematics of the elementary processes. B atoms are deposited with a flux $F$. Once on the surface, they
can diffuse with diffusion constant $D_{\rm B}$, aggregate to an island, nucleate one, or exchange vertically with a underlying 
A atom of the substrate at an effective rate $r$. Emerging A atoms can diffuse with diffusion constant $D_{\rm A}$, 
aggregate to an island or form a new one, but cannot exchange back. Every island, composed of one or two types of atoms, 
occupies a single lattice site and grows by aggregation. Atom detachment and evaporation are not allowed.}\label{fig0}
\end{figure}

A substrate, which consists of A atoms, is represented by a square lattice of 
$L\times L$ sites, with periodic boundary conditions to avoid edge effects. On 
this lattice, we deposit  B atoms, which perform random walks and 
undergo place exchange with substrate atoms by a phenomenological constant 
rate $r$. When any two diffusing atoms meet, they form a point 
island. Theses islands do not diffuse nor break, and grow irreversibly by 
aggregation of other atoms.   Every island occupies
only one lattice site, in spite of the number of atoms that compose it. 
Detachment and evaporation are not considered.
Structures result of mixed composition because two kind of atoms 
are involved. During time evolution, we take into account the following 
processes (shown schematically in 
figure \ref{fig0}):\par

(a) Deposition: starting from an initially flat substrate consisting of A 
atoms, each empty site of the lattice is occupied by an B atom  with 
probability per unit time $F$. Every simulation  runs until the number 
of atoms deposited per site reaches a desired value $\Theta$.\par

(b) Intermixing: when a diffusing B atom (not bounded to an island) lays on a A atom of the substrate, the former
exchanges with the latter with rate $r$. After an exchange, the B atom remains irreversibly 
incorporated to the substrate (no exchange back) and the A atom starts diffusing.\par 

(c) Diffusion: any  unbounded A (B) atom on the surface diffuses with 
diffusion constant $D_{\rm A}$ ($D_{\rm B}$), by hopping among 
nearest-neighbors lattice sites.\par

(d) Nucleation: when any two diffusing atoms (either A or B) meet, they form a stable non-moving island. Each island 
acts as a nucleation center and occupies only one site on the lattice.\par 

(d) Aggregation: when an diffusing atom, regardless of its type, hops to a 
site occupied by an island, the former aggregates to the latter, which 
increases its number of particles by one. Detachment events are not 
allowed, i. e. islands grow irreversibly. \par

%%%%%%%%%%%%%%%%%%%%%%%%%%%%%%%%%% FIGURE 0 %%%%%%%%%%%%%%%%%%%%%%%%%%%%%%%%%%%%%%
%%%%%%%%%%%%%%%%%%%%%%%%%%%%%%%%%%%%%%%%%%%%%%%%%%%%%%%%%%%%%%%%%%%%%%%%%%%%%%%%%%%

 We are interested in island formation at low enough densities, 
at which the lattice mismatch 
and most of the interactions among diffusing atoms can be neglected. 
We expect that, in early stages of growth, this point island model is 
useful to describe different properties of 
the system, such as island density and interface composition. 
We perform simulations with $\Theta$ always below $0.2$ mono layer
(ML). \par

According to the above described processes, the model dynamics depend on four 
parameters: 
the deposition flux $F$,  exchange rate $r$, and the 
diffusion constants $D_{\rm B}$ and $D_{\rm A}$.
However, at a given coverage $\Theta$ (or time $t=\Theta/F$), the 
surface structure is determined by only three non-dimensional numbers, which 
are the ratios $ \epsilon=a^4F /D_{\rm B}$, 
$\kappa =D_{\rm A}/D_{\rm B}$ and 
$\pi=a^2r/D_{\rm B}$, where $a$ is the lattice 
constant (in the following we set $a=1$). 
In this work, we will show results for $\kappa \le 1$ , though it is easy to 
extend them to other values of $\kappa$.
Note that this model reduces to the standard point islands one, when all atoms 
diffuse with the same constant, i., e., for $\kappa=1$ \cite{6PRE}.
In the following, we study the density and composition of the islands as a 
function of $ \epsilon$, for different values of the non-dimensional 
intermixing  and diffusion ratios $\pi$ and $\kappa$, 
respectively .

\section{Results}
\label{discussion}

\subsection{Island density}
\label{Island density}

\begin{figure}[!ht]
\begin{center}
\includegraphics[width=.75\linewidth]{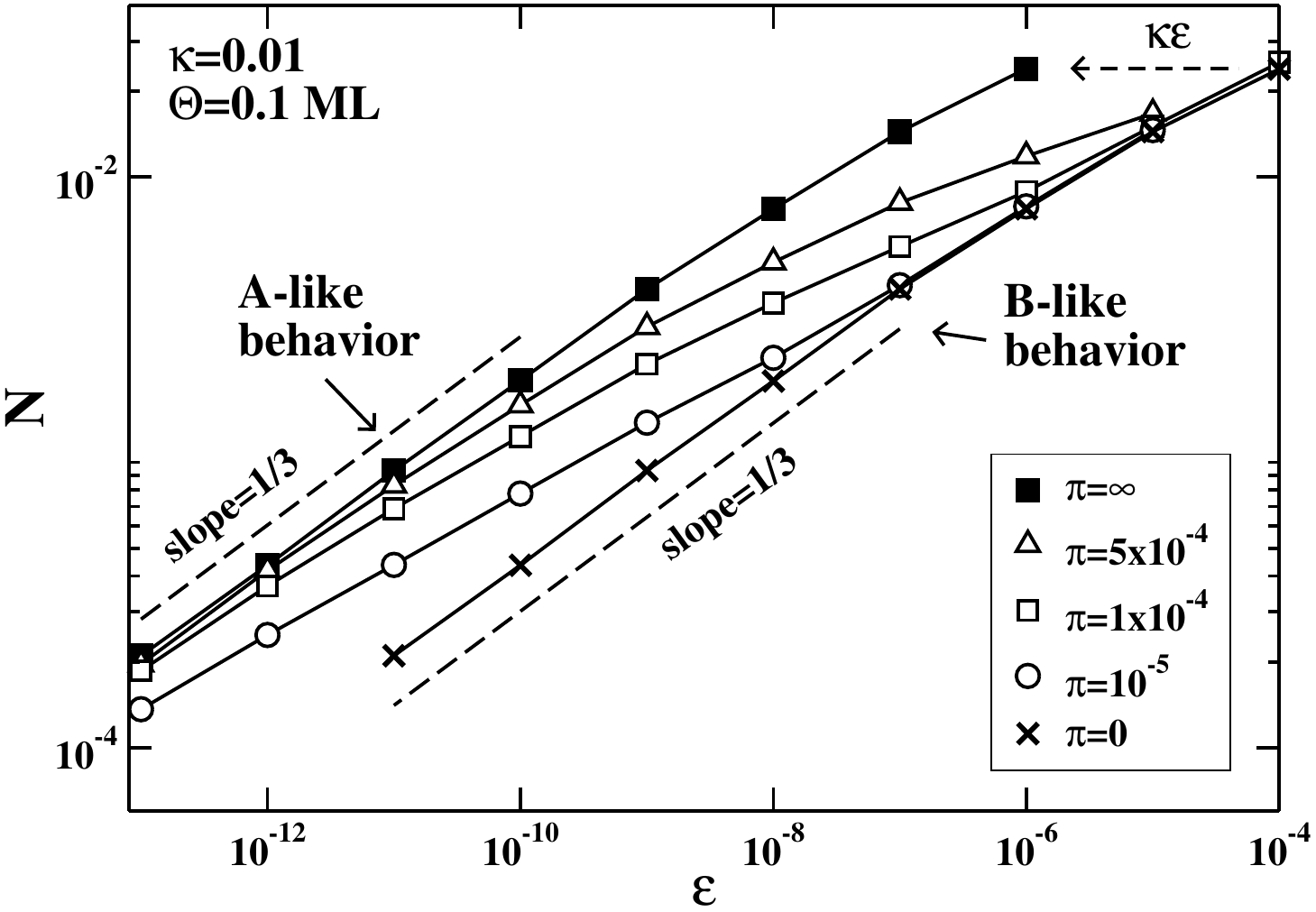}
\end{center}
\caption{The island density against the non-dimensional incoming flux in log-log scales, for fixed diffusion rate $\kappa$ and 
coverage $\Theta$, and different values of the non-dimensional intermixing $\pi$. When $\pi=0$, 
diffusing atoms are B (crosses), while  for $\pi = \infty$, most of the B atoms are incorporated to the substrate,  
and exchanged A atoms  move with a diffusion constant $D_{\rm A} = k D_{\rm B}$. A  data collapse of the solid squares and crosses can be 
obtained through the scaling $ \epsilon \to \kappa \epsilon$. Note that, given $\kappa<1$ and $\pi$,  
we observe an  A-like behavior when $ \epsilon$ is low enough,  and a B-like behavior when $ \epsilon$ is large enough.}
\label{fig1}
\end{figure}

In this part, we analyze the island density $N$ as a function of model 
parameters. Surface composition will be addressed in 
section \ref{Surface composition}.
At a given temperature $T$, which determines the diffusion 
constants of atoms, the number of islands depends on $ \epsilon$, a 
measure of the relationship between deposition and diffusion of B atoms. 
As $ \epsilon$ increases each diffusing atom performs a lower number of 
hops in the mean time between incoming particles. 
This leads to a higher density of monomers, and to a greater nucleation 
probability. Thus, the island density increases with $ \epsilon$. It is 
known that, at a fixed coverage, the average number of island per lattice site 
$N$ behaves as $N \sim  \epsilon^{\chi}$, for $ \epsilon$ small 
enough.  The exponent $\chi$ 
depends on the effective dimensionality of diffusion. For the two-dimensional 
case, $\chi=\frac{1}{3}$~\cite{6PRE, 7PRE, 8PRE, 9PRE}.
Examples of the behavior the island density, obtained numerically,  as a 
function of the non-dimensional incoming flux
are shown in figure \ref{fig1} for $\kappa=0.01$, $\Theta=0.1$ML, and 
$\pi=\infty, 5\times 10^{-4}, 1\times 10^{-4}, 
1\times 10^{-5}$, and $0$. The two-dimensional exponent $\chi=\frac{1}{3}$ is 
in agreement with these results for $ \epsilon$
small enough.

%%%%%%%%%%%%%%%%%%%%%%%%%%%%%%%%%% FIGURE 1 %%%%%%%%%%%%%%%%%%%%%%%%%%%%%%%%%%%%%%
%%%%%%%%%%%%%%%%%%%%%%%%%%%%%%%%%%%%%%%%%%%%%%%%%%%%%%%%%%%%%%%%%%%%%%%%%%%%%%%%%%%

To go beyond the slow deposition regime, in what follows we discuss the dependence of $N$ on $\pi$, 
for intermediate values of $ \epsilon$.
A simple situation corresponds to $\pi=0$, when no intermixing takes place and the model reduces to  the standard point
islands model. Note that, with respect to island density, this condition is equivalent to $\kappa=1$ (and any $\pi$),
which means that all particles diffuse in the same manner, with diffusion constant $D_{\rm B}$ (B-like behavior).  
Other simple situation  occurs when $\pi=\infty$. In this case, each entering B atom exchanges instantaneously 
with the first A atom it lays on. Thus, diffusing atoms all come from the substrate, and island dynamics are governed by the 
diffusion constant $D_{\rm A}$ (A-like behavior). These limit behaviors are summarized in Table~\ref{table}. 
 Let us remark that the function $N( \epsilon )$ for 
$\pi=\infty$ can be obtained from that for 
$\pi=0$ by rescaling $ \epsilon \to \kappa \epsilon$. We return to this point in section \ref{Mean-field}.\par

%%%%%%%%%%%%%%%%%%%%%%%%%%%%%%%%% table   %%%%%%%%%%%%%%%%%%%%%%%%%%%%%%%%%%%%%%
\begin{table}
\caption{Limit behaviors. Island dynamics are governed by diffusion of either 
A or B atoms for extremely large or small values of $\pi$ or 
$ \epsilon$.  For the special case in which both species diffuse in the 
same manner, the model reduces to the standard point island model
(last row)  \cite{6PRE}.}
\vspace{1.5 cm}
\begin{indented}
\item[]\begin{tabular}{@{}lccc}
\br
 $\kappa\hspace*{1cm}$ &$\pi$& $ \epsilon$	
& Dynamics governed by\\ 
\mr
& $0$		& any	&\\
&	&		& $D_{\rm B}$\\
& any	&large enough 	&\\
$\neq 1$ &		&\\ \cline{2-4}  
&$\infty$	&any	&\\
&		&		&$D_{\rm A}$\\
&any	&small enough	&\\ 
\mr				
$1$ &any	& any	&$D_{\rm A}=D_{\rm B}$\\
\br
\end{tabular}
\end{indented}
\label{table}
\end{table}
%%%%%%%%%%%%%%%%%%%%%%%%%%%%%%%%%%%%%%%%%%%%%%%%%%%%%%%%%%%%%%%%%%%%%%%%%%%%%%%%

It is interesting to note that the A-like and B-like regimes can also be 
observed for other values of $\pi$, by tuning the parameter 
$ \epsilon$. For instance, as $ \epsilon$ increases, both the 
nucleation and aggregation mean times ($t_{\rm n}$ and 
$t_{\rm a}$, respectively) decrease. For large enough $ \epsilon$, they 
become much shorter than the intermixing mean time $r^{-1}$, and 
island dynamics are governed by diffusing B atoms, which have a little 
exchange probability. In contrast, for low enough $ \epsilon$, $t_{\rm n}$ and $t_{\rm a}$ are much longer than $r^{-1}$ and most of the 
moving atoms are of kind A. The presence of atoms from the substrate forming 
part of islands has been observed in experiments carried out at
high temperatures, which corresponds to the second situation. For instance, 
the growth of Nb on Fe(110) and Fe on Nb(110) form surface alloy at 
temperatures above $800$ K and a sufficient epitaxial quality of layer by layer
 can be obtained without intermixing of Nb and Fe, at room 
temperature~\cite{Wolf06}, the growth of Au on Fe(001) exhibits alloy at 
temperatures higher than $370$ K~\cite{Kempen01}. 
Estimations of the characteristic times $t_{\rm n}$ and $t_{\rm a}$ are given 
in section \ref{Scaling}, where a scaling form of island density is obtained 
using mean-field approximations.

\subsection{Surface composition}
\label{Surface composition}

\begin{figure}[!ht]
\begin{center}
\includegraphics[width= .75\linewidth]{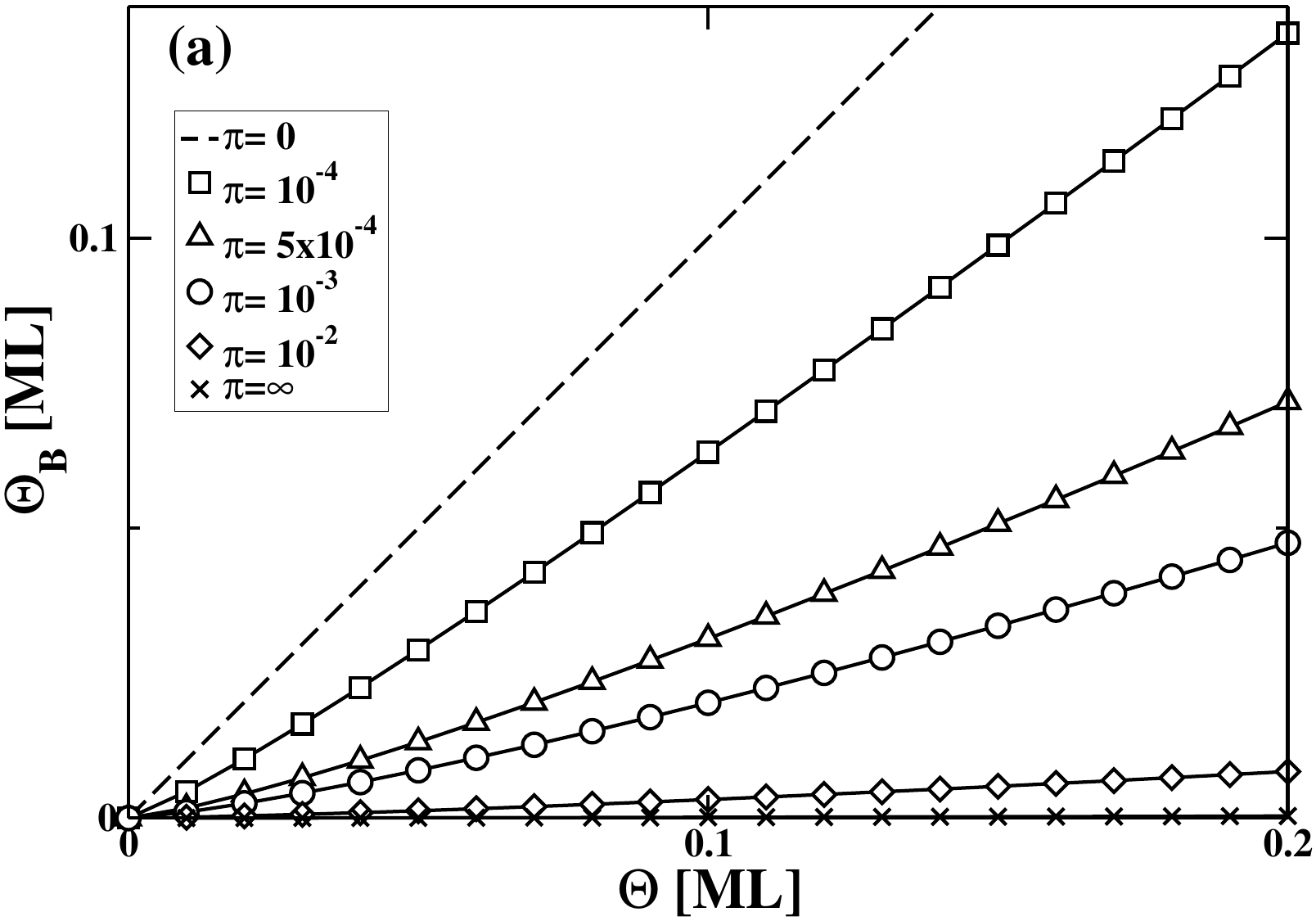}
\includegraphics[width= .75\linewidth]{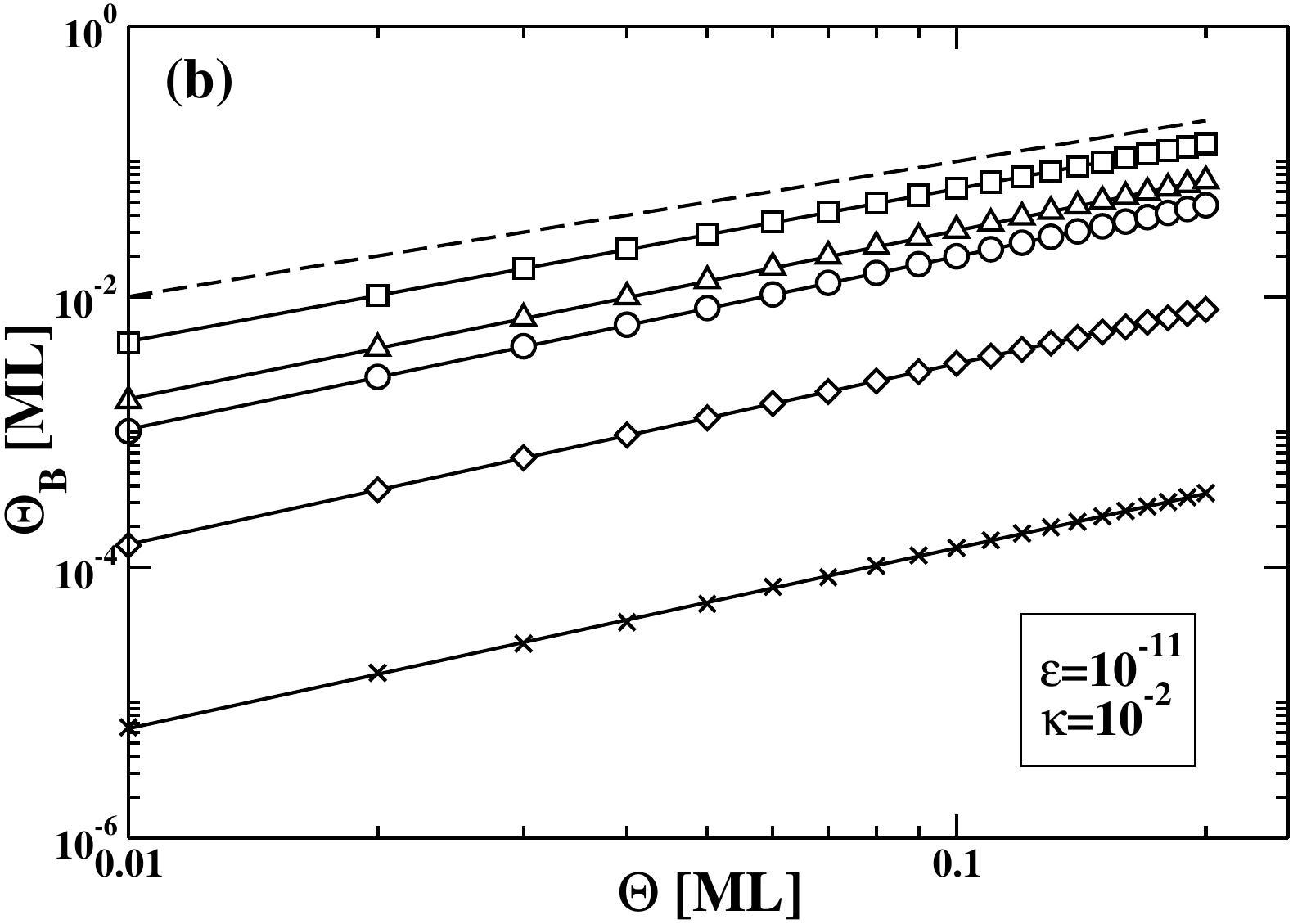}
\end{center}
\caption{(a) Density of B atoms that stay on the substrate $\Theta_{\rm B}$, as 
a function of the coverage $\Theta$, for $\kappa=10^{-2}$ 
and $ \epsilon=10^{-11}$ and  different values of $\pi$.  
$\Theta_{\rm B}$ decreases with $\pi$, as expected. Except for 
$\pi=0$, the curves are concave upward.
(b) Same plots in log-log scale. The slopes are, from top to bottom  
$\beta = 1, 1.116, 1.231, 1.266, 1.329, 1.335$. 
}\label{3}
\end{figure}

The amount of B atoms incorporated to islands per site $\Theta_{\rm B}$ should 
decrease with the increasing of the intermixing rate.
This is clearly observed in figure \ref{3}(a), where we show the behavior 
of $\Theta_{\rm B}$ as a function of $\Theta$ for 
$ \epsilon=10^{-11}$, $\kappa=0.01$, and 
$\pi=\infty,\;1\times 10^{-2},\;1\times 10^{-3},\;
5\times 10^{-4}$ 
and $1\times 10^{-4}$. Every set of data points fits with a curve concave 
upward, i., e., its derivative is monotonically 
increasing, which originates in the fact that, as the island density is a 
growing function of $\Theta$, the larger
the coverage, the higher the aggregation probability for diffusing B atoms 
before they intermix with A atoms. 
Note that, at a given coverage, the concavity increases with 
$\pi$, due to the increasing of the 
intermixing/aggregation ratio.
In figure \ref{3}(b)  we show the same plots in log-log scale, and the measured 
effective exponents (greater than 1)
for the each value of $\pi$.\par

                                                                              %%%%%%%%%%%%%%%%%%%%%%%%%%%%%%%%%% FIGURE 3 %%%%%%%%%%%%%%%%%%%%%%%%%%%%%%%%%%%%%%
%%%%%%%%%%%%%%%%%%%%%%%%%%%%%%%%%%%%%%%%%%%%%%%%%%%%%%%%%%%%%%%%%%%%%%%%%%%%%%%%%%%

\begin{figure}[!ht]
\begin{center}
\includegraphics[width= .75\linewidth]{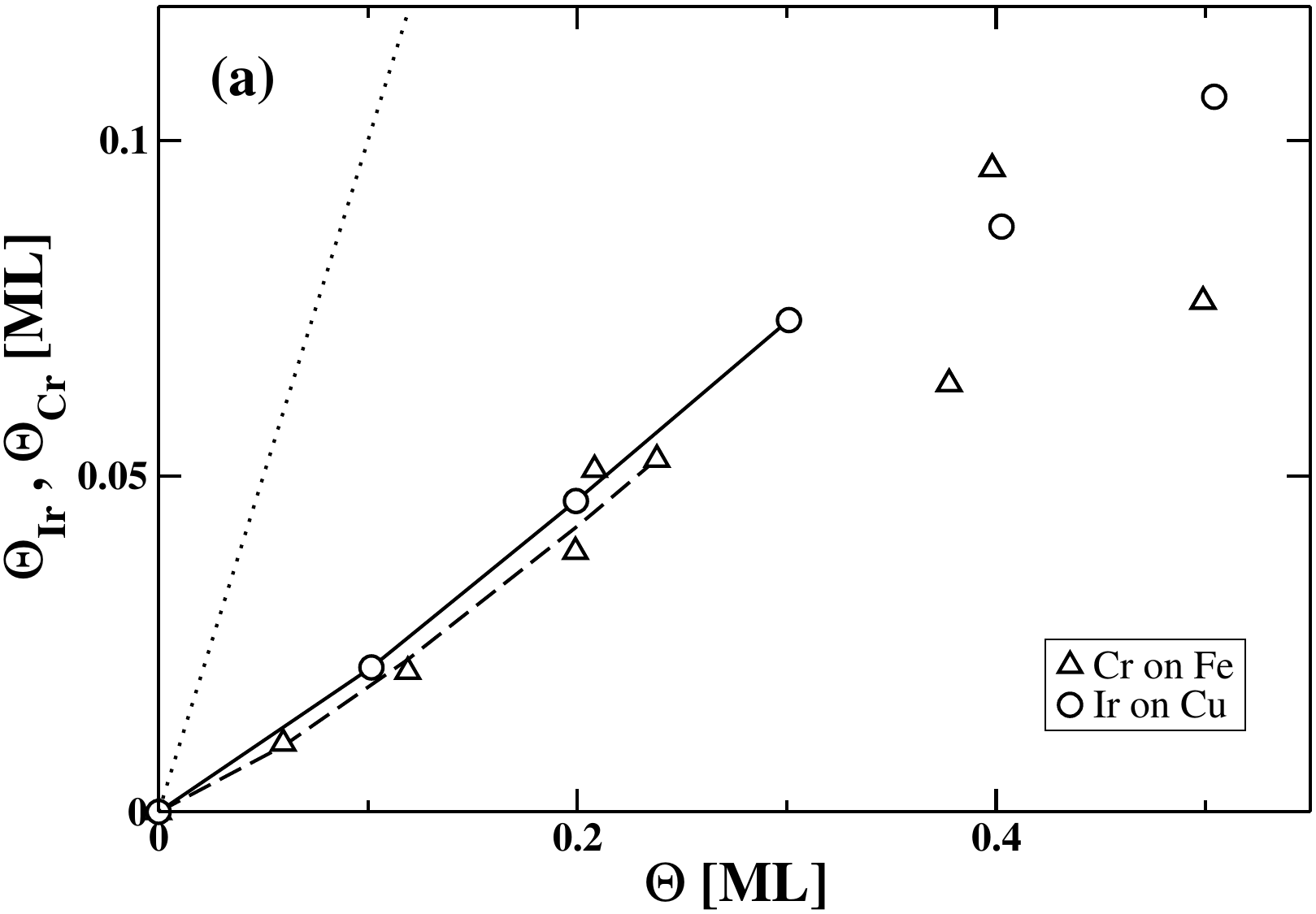}
\includegraphics[width= .75\linewidth]{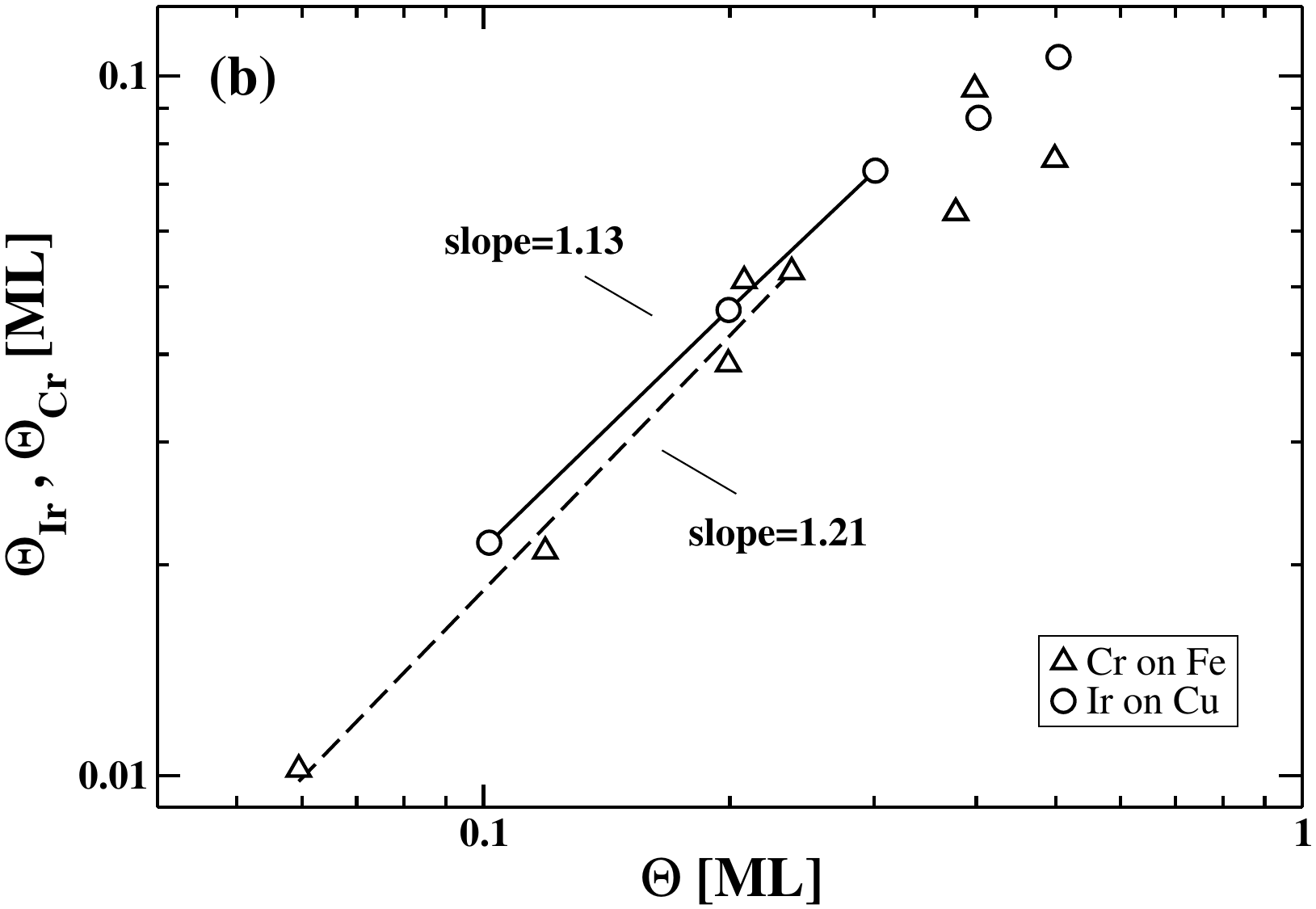}
\end{center}
\caption{(a) Experimental density of deposited atoms staying atop as function 
of the coverage, for Ir on  Cu(100) (circles) 
and Cr on Fe(001) (triangles). The working temperature was  $200$ K for the 
former, and $300$ K for the 
latter~\cite{Niehus99, Celotta96}. 
The change in the concavity observed for $\Theta$ in the interval 
($ 0.2-0.3 \mbox{ML}$) indicates a crossover value of the 
coverage, at which the assembling of 3D islands starts. The dotted line 
stands for the layer-by-layer growth without intermixing. 
Cr concentrations  were measured on exposed regions of the substrate, but no 
significant difference has been 
found when taking into account Cr concentrations on islands, in this range of 
coverages \cite{Celotta96}.
(b) Same plots in log-log scale. Effective exponents greater than 1 are 
obtained at low coverages.
}
\label{exp}
\end{figure}

The upward concavity of $\Theta_{\rm B}$ as a function of $\Theta$, 
has been observed in experiments at low coverages. This is, for example, the 
case of  epitaxial growth of Ir on Cu and Cr on Fe~\cite{Niehus99, Celotta96}. 
For Ir (Cr) atoms, it seems energetically more favorable to be embedded via 
place exchange in the Cu (Fe) substrate rather than staying atop, which 
redounds in intermixing. Experimental data of the amount of 
Ir (Cr) atoms that stays on the surface as a function of coverage (extracted 
from \cite{Niehus99, Celotta96}) are shown 
in figure \ref{exp}(a). At low enough coverage, the experimental data fit to 
curves conclaves upward. Thus, effective exponents greater 
than one are measured for $\Theta_{\rm{Ir (Cr)}}$ versus $\Theta$, at low 
coverage (see figure \ref{exp}(b)). Similar behaviors 
were observed for Fe on GaAs \cite{Ionescu05} and Cu on Ir \cite{Niehus00}.\par

%%%%%%%%%%%%%%%%%%%%%%%%%%%%%%%% FIGURE 4 %%%%%%%%%%%%%%%%%%%%%%%%%%%%%%%%%%%%%%
%%%%%%%%%%%%%%%%%%%%%%%%%%%%%%%%%%%%%%%%%%%%%%%%%%%%%%%%%%%%%%%%%%%%%%%%%%%%%%%%

The change of concavity detected in experiments at intermediate values of 
$\Theta$ ($\simeq 0.2-0.3 \mbox{ML}$ in figure \ref{exp})
can be explained by the onset of 3D island growth or by an intermixing rate 
growing with $\Theta$ \cite{Niehus00, Winter96, Winter97}. According to the 
model studied in this work, this effect cannot be 
attributed to intermixing, if its rate does not depend on the coverage; 
even for large values of $\pi$.
As discussed above, and shown in figure \ref{3}(b) 
($\Theta_{\rm B}\sim\Theta^\beta$ 
with $\beta >1$ for $\pi\neq 0$), 
the derivative of $\Theta_{\rm B}$ always increases with $\Theta$. 
In the way to minimize the free energy, the atomic structures can reduce their 
surface by assembling 3D islands, and the deposited
B atoms can intermix with substrate A atoms. At low enough coverages, the 
latter is the most relevant process. 
As coverage increases, the configuration that minimizes the surface free 
energy most likely involves 3D islands. The crossover 
between both behaviors will depend on the particular reactants. Although 
interesting, the study of this crossover is beyond 
the scope of our model.

\subsection{Mean-field evolution}
 \label{Mean-field}

From the rules described in section~\ref{model}, the rate evolution 
equations for the total monomer and island densities, at low enough 
coverages, and using mean-field arguments are
\begin{eqnarray}
        \frac{\rmd\left( n_{\rm A}+n_{\rm B}\right)}{\rmd t} =& F - 
\left[ k_{\rm A} n_{\rm A}^2 + k_{\rm B} n_{\rm B}^2 + \left(k_{\rm A} + 
k_{\rm B}\right)n_{\rm A}n_{\rm B} \right] \nonumber\\
&-\left( k_{\rm A} n_{\rm A} + k_{\rm B} n_{\rm B} \right)N
\label{rate1}
\end{eqnarray}
\begin{eqnarray}
&\frac{\rmd N}{\rm d t}& = F\left( n_{\rm A} + n_{\rm B} \right) +  
\left[ k_{\rm A} n_{\rm A}^2 + k_{\rm B} n_{\rm B}^2 + \left(k_{\rm A} 
+ k_{\rm B}\right)n_{\rm A}n_{\rm B} \right]\;{,}
\label{rate2}
\end{eqnarray}

\noindent
where $n_{\rm A}$ ($n_{\rm B}$) is the A (B) monomer density and $k_{\rm A}$ 
($k_{\rm B}$) governs the A (B) monomer attachment rate 
(it is known that $k_{\rm A}\sim D_{\rm A}$ and $k_{\rm B}\sim D_{\rm B}$, 
for point islands \cite{9PRE}).\par

The first term in the right-hand side of (\ref{rate1}) corresponds to the 
increase of monomers due to the deposition of B atoms.
The  second and the last, to its decrease, due to nucleation and aggregation 
to islands, respectively.
Note that the  parameter $r$ does not appear in (\ref{rate1}).
This equation refers to total monomer 
density variation, which is no affected by the intermixing. 
In the right-hand side of (\ref{rate2}), both terms stand for nucleation. 
The first, that which occurs when a B atom
is deposited on a diffusing monomer, the second corresponds to nucleation 
by diffusion. As islands cannot break, they always increase in number with 
time.\par

We can rewrite the rate equations in terms of the coverage 
$\Theta=F t$ (rather than time $t$) as

 \begin{eqnarray}
 &\frac{\rmd\left( n_{\rm A}+n_{\rm B}\right)}{\rmd\Theta}& = 
1 - \frac{\left[ \kappa n_{\rm A}^2 + n_{\rm B}^2 + \left(\kappa + 1\right)
n_{\rm A}n_{\rm B} \right] }{ \epsilon }- \frac{\left( \kappa n_{\rm A} +  n_{\rm B} \right) N}{ \epsilon }\label{rate3}\\
&\frac{\rmd N}{\rmd\Theta}& = \left( n_{\rm A} + n_{\rm B} \right) + 
\frac{\left[ \kappa n_{\rm A}^2 + n_{\rm B}^2 + \left(\kappa + 1\right)
n_{\rm A}n_{\rm B} \right] }{ \epsilon }\;{.}
\label{rate4}
\end{eqnarray}
For small enough $ \epsilon$, a quasi-stationary regime exists, in which 
$\left(n_{\rm B} + n_{\rm A}\right)\ll N \ll 1$,
and ${\rm d}\left( n_{\rm A} + n_{\rm B}\right)/\rm d\Theta \cong 0$. 
In addition, in this regime $n_{\rm B}\ll n_{\rm A}$, 
provided that $\pi \neq 0$. Thus, by retaining only the leading terms in (\ref{rate3}) and (\ref{rate4}), we get 
$n_{\rm A} \sim  \epsilon/\kappa N$ and 
$dN/d\Theta \sim \kappa n_{\rm A}^2/ \epsilon$, 
which lead to
\begin{equation}
N\sim \left( \frac{\Theta  \epsilon }{\kappa } \right)^{1/3} \mbox{\;\;.}
\label{chi}
\end{equation}
This expression holds, for $\kappa\neq 0$ and $\pi \neq 0$, 
at small enough $ \epsilon$, as confirmed by the
results of simulations in figure \ref{fig1}.

Regarding the quantity $\Theta_{\rm B}$ as a function of $\Theta$, it is easy 
to obtain the exponents related to its power-law behavior
in the limits $ \pi \to \infty $ and 
$\pi \to 0$. In the first case, for $ \epsilon$ small 
enough, B atoms stay atop only if they are deposited directly on islands, and 
then  $\rmd \Theta_{\rm B}/\rmd t \cong F N$, which, using 
(\ref{chi}), gives
\begin{eqnarray}
\Theta_{\rm B} \sim \Theta^{4/3} \left( \epsilon /\kappa \right)^{1/3} 
\mbox{\;\;.}
\label{coverage}
\end{eqnarray}
In contrast, when $\pi=0$, all diffusing atoms are B and 
$\Theta_{\rm B} = \Theta$. These 
limit behaviors are confirmed by simulations, as shown in figure \ref{3}. 
We can observe in the same 
figure (part (b)) that, for intermediate values of $ \pi$, 
and $\Theta$ in the range 
$[0.01\mbox{ML}-0.2\mbox{ML}]$,  
$\Theta_{\rm B} \sim \Theta^\beta$; with an effective exponent $\beta$ that 
decreases from $\frac{4}{3}$ to 
$1$, when $\pi$ moves from $\infty$ to $0$. 

\subsection{Scaling of the island density}
\label{Scaling}
As discussed in section \ref{Island density}, for low coverage, 
and fixed $\pi$ ($\neq 0$ or $\infty$) and $\kappa$ ($\neq 0$),
 island dynamics are governed by the diffusion of A (B) atoms for small (large) 
enough $ \epsilon$ (see Table I). 
The extensions of the A-like and B-like regimes in the parameter space 
depends on the involved 
characteristic times $t_{\rm n}$ and $t_{\rm a}$.
The nucleation time of an island composed of a pair of B atoms can be 
estimated by considering that, in an average time $t_{\rm n}$, a B atom is 
deposited in one of the mean number of distinct sites visited by a diffusing B 
atom $S(t_{\rm n})$, i., e., $F S(t_{\rm n}) t_{\rm n} \sim 1$. As, for 
two-dimensional diffusion, 
$S(t_{\rm n}) \sim D_{\rm B} t_{\rm n}$~\cite{Stanley92}, we arrive to the 
expression 
\begin{eqnarray}\label{tn}
  t_{\rm n} \sim \frac{1}{D_{\rm B}  \epsilon ^{1/2}} \;{.} 
\end{eqnarray}
To estimate the aggregation time of a diffusing B atom, we assume
that the mean number of distinct sites visited by a B atom in this time
$S(t_{\rm a})$ is proportional to the average number of empty sites per island,
i., e., $S(t_{\rm a}) \sim 1/N$. Then, taking into account~(\ref{chi}), and 
the above mentioned behavior of $S(t_{\rm a})$, the estimates results
\begin{eqnarray}\label{ta}
 t_{\rm a} \sim \frac{1}{D_{\rm B}}\left( \frac{\kappa }{\Theta   \epsilon}\right) ^{1/3}\;{.} 
\end{eqnarray}
Note that $t_{\rm a} < t_{\rm n}$, for $ \epsilon$ small enough.\par

When both $t_{\rm a}$ and $t_{\rm n}$ are much longer than the intermixing 
time, i., e., \mbox{$t_{\rm n}>t_{\rm a}\gg r^{-1}$}, most of the 
deposited B atoms intermix with the substrate, and the diffusing atoms are 
predominantly A. 
According to (\ref{tn}), the A-like behavior occurs for 
$ \epsilon\ll \epsilon_{\rm A}$, where
\begin{eqnarray}\label{A-regime}
  \epsilon_{\rm A} \sim \frac{\kappa \pi^3}{\Theta}\;\mbox{.}
\end{eqnarray}

%%%%%%%%%%%%%%%%%%%%%%%%%%%%%%%%%% FIGURE col %%%%%%%%%%%%%%%%%%%%%%%%%%%%%%%%%%%%%%
\begin{figure}[!ht]
\begin{center}
\includegraphics[width= .75\linewidth]{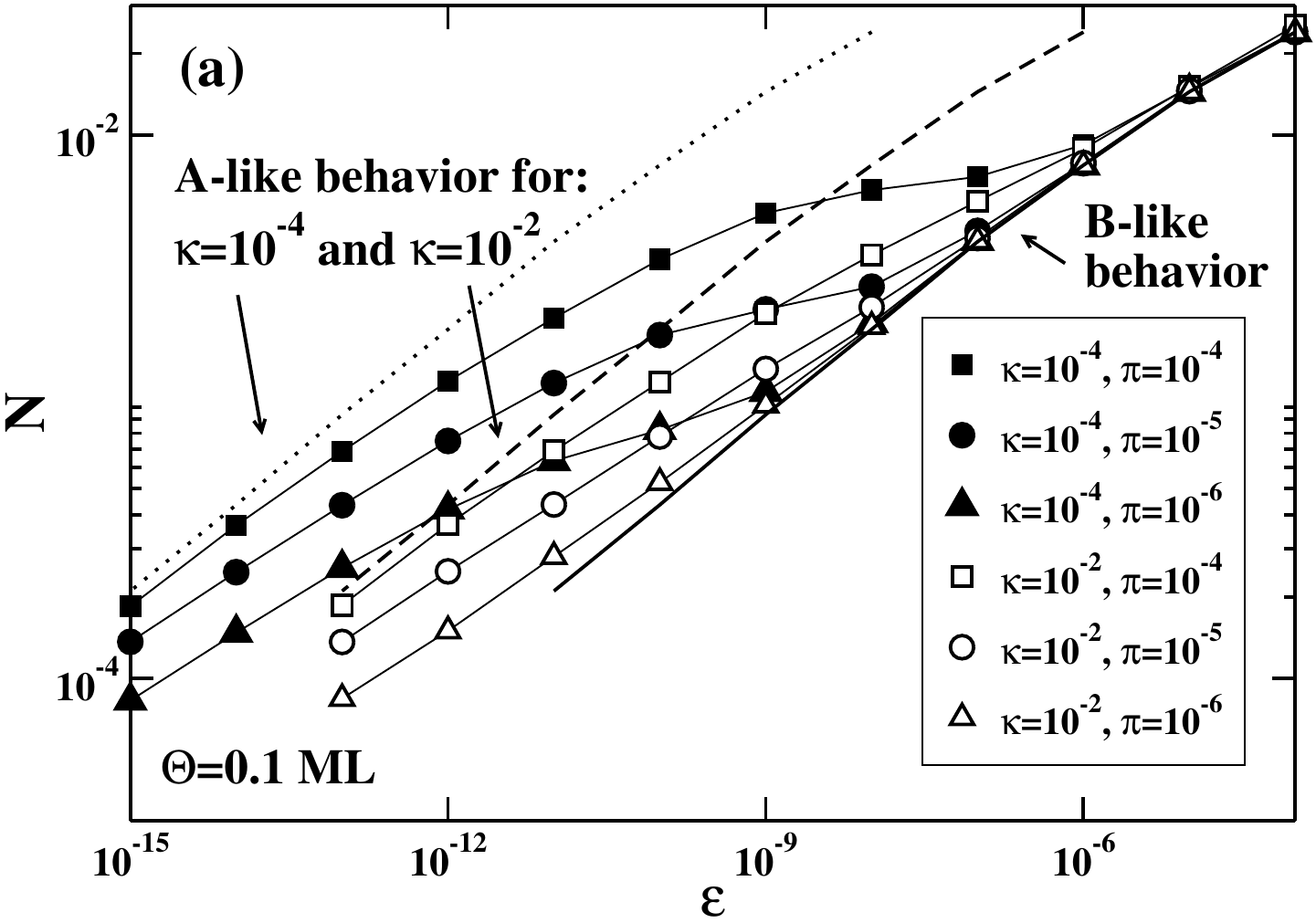}
\vspace{.5cm}\\
\includegraphics[width= .75\linewidth]{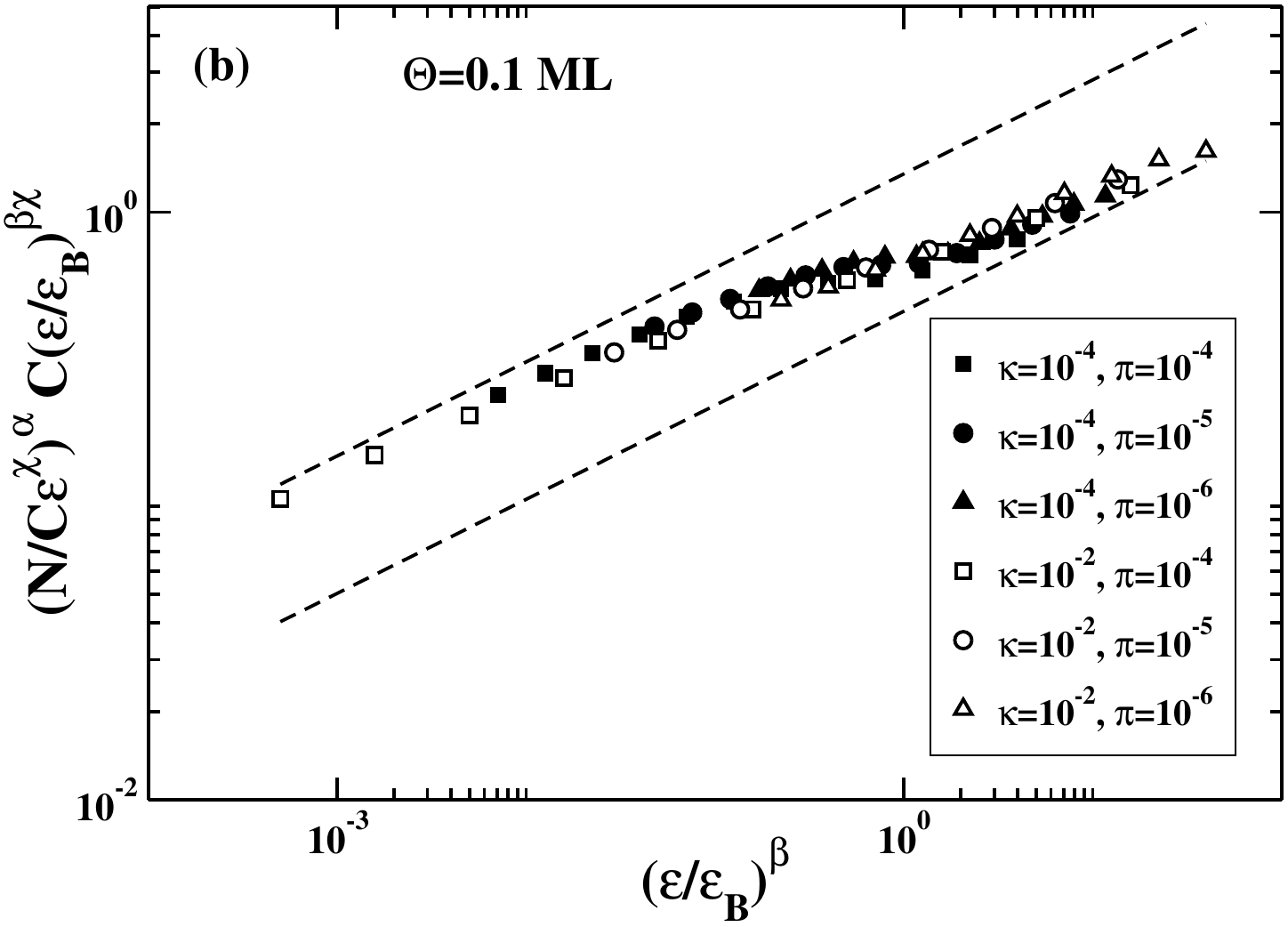}
\end{center}
\caption{(a) The island density against the non-dimensional incoming flux, in 
log-log scales, for different values of $\pi$ and $\kappa$. 
The dotted and dashed lines correspond to A-like behaviors 
($\pi=\infty$); $\kappa =10^{-4}$ for the former,  $\kappa=10^{-2}$ for the latter. The solid line corresponds to $\pi=0$. 
When $ \epsilon\ll \epsilon_{\rm A}$ (see (\ref{A-regime})), the data 
points for $\kappa=10^{-4}$ (solid symbols) and $\kappa=10^{-2}$ (open symbols)
approach the dotted and dashed curves, respectively. The B-like
behavior occurs when $ \epsilon_{\rm B}\ll \epsilon$ 
(see (\ref{B-regime})).
(b) Scaling form of the island density for the same data in part (a): 
$\alpha=-\log(\kappa), \beta=\log(\epsilon_B/\epsilon_A)$, 
and the dashed lines represent  A-like (upper) and  B-like (lower) behaviors; 
see the main text for further details}\label{colapso}
\end{figure}
%%%%%%%%%%%%%%%%%%%%%%%%%%%%%%%%%%%%%%%%%%%%%%%%%%%%%%%%%%%%%%%%%%%%%%%%%%%%%%%%%%%

In contrast, most of diffusing atoms are B when 
$t_{\rm a} < t_{\rm n} \ll r^{-1}$.
Thus, from (\ref{tn}), the B-like regime occurs when 
$ \epsilon_{\rm B}\ll \epsilon$, where 
\begin{eqnarray}\label{B-regime}
  \epsilon_{\rm B} \sim \pi^2\;\mbox{.}
\end{eqnarray} 

The crossover scales given by (\ref{A-regime}) and (\ref{B-regime})
allow to collapse the curves corresponding to $N$ as a function of 
$ \epsilon$, for different values of $\kappa$ and $\pi$, 
provided that $\pi$ is small enough. 
In figure \ref{colapso}(a) we have plotted this function using the results
of numerical simulations for $\Theta=0.1$ML, $\kappa=10^{-2}$ and $10^{-4}$, 
and $\pi=10^{-4},\;10^{-5}$ and $10^{-6}$. 
{  Note that the behavior of $N(\epsilon)$ (for fixed $\theta, \kappa$,
and $\pi$) can be expressed as
$N(\epsilon)=C \epsilon^\chi G(\epsilon)$. In this equation,
$C \epsilon^\chi$ stands for the B-like behavior, while the function
$G(\epsilon)$ takes a constant value $\sim-\log(\kappa) $ for $\epsilon\ll\epsilon_A$, and $1$ for $\epsilon_B\ll\epsilon$; decreasing monotonically between $\epsilon_A$ and $\epsilon_B$.  
}
Since two different crossover  exist,  at a given $\Theta$ the data
collapse is achieved in two steps. First, every curve { 
corresponding to $G(\epsilon)$ in figure 
\ref{colapso}(a) is rigidly translated to move the second crossover point 
to the origin, by plotting $(N/C\epsilon^\chi)$ as a function 
of $\epsilon/\epsilon_{\rm B}$. Then, the $y$ axis is rescaled by 
$\alpha=(-\log(\kappa))^{-1}$ and the $x$ axis by $\beta=(\log( \epsilon_{\rm B}/  \epsilon_{\rm A}))^{-1}$. 
A last transformation (a backwards rotation $y \rightarrow y\; Cx^\chi$) is included
in order to recover the overall behavior of $N$.
}
The finally resulting plot, for the data in figure \ref{colapso}(a), is shown 
in the part (b) of the same figure. 
The very good collapse on a single curve is apparent, and gives support to
the idea of universality, according to which, at low coverages, the island 
density satisfies
{ 
\begin{equation}
{\log\left(\displaystyle\frac{N(\epsilon)}{C\epsilon^\chi}\right)}=
-\log(\kappa)
{\cal G}\left[
{\displaystyle\log\left(\frac{ \epsilon}{ \epsilon_{\rm B}}\right)\over
\displaystyle\log\left(\frac{ \epsilon_{\rm B}}{ \epsilon_{\rm A}}\right)
}
\right]\;\mbox{,}
\end{equation}
}
where ${\cal G}(x)$ is a universal scaling function.

\subsection{Dynamic scaling of the island-size distribution}
\label{Dynamic Scaling}

An important quantity in the description of island growth, is the 
size distribution function $n_{\rm s}(\Theta)$, which gives the number per 
site of islands of size $s$ (composed of $s$ atoms), 
at a coverage $\Theta$. It is well established 
\cite{6PRE,Family:1984,Amar,Family:1988,Feldman} 
that, {  in the case of the standard  irreversible aggregation 
model, where all particles diffuse with same constant of diffusion $D$,
for low enough values of the ratio $F/D$, the low-coverage dynamics are 
self-similar and the island size distribution is described by 
\begin{eqnarray}\label{scaling}
n_{\rm s} (\Theta ) = \frac{\Theta}{\left < s \right >^2} f \left( \frac{s}{\left < s \right >}\right)
\end{eqnarray}
where the form of the scaling function $f(x)$ is universal, in the sense that 
it does not depend upon the details of the model, such as the lattice type and 
the coordination number, but rather depends on more global variables. 

\begin{figure}[!h]
\begin{minipage}[b]{0.5\linewidth}
\includegraphics[width=\linewidth]{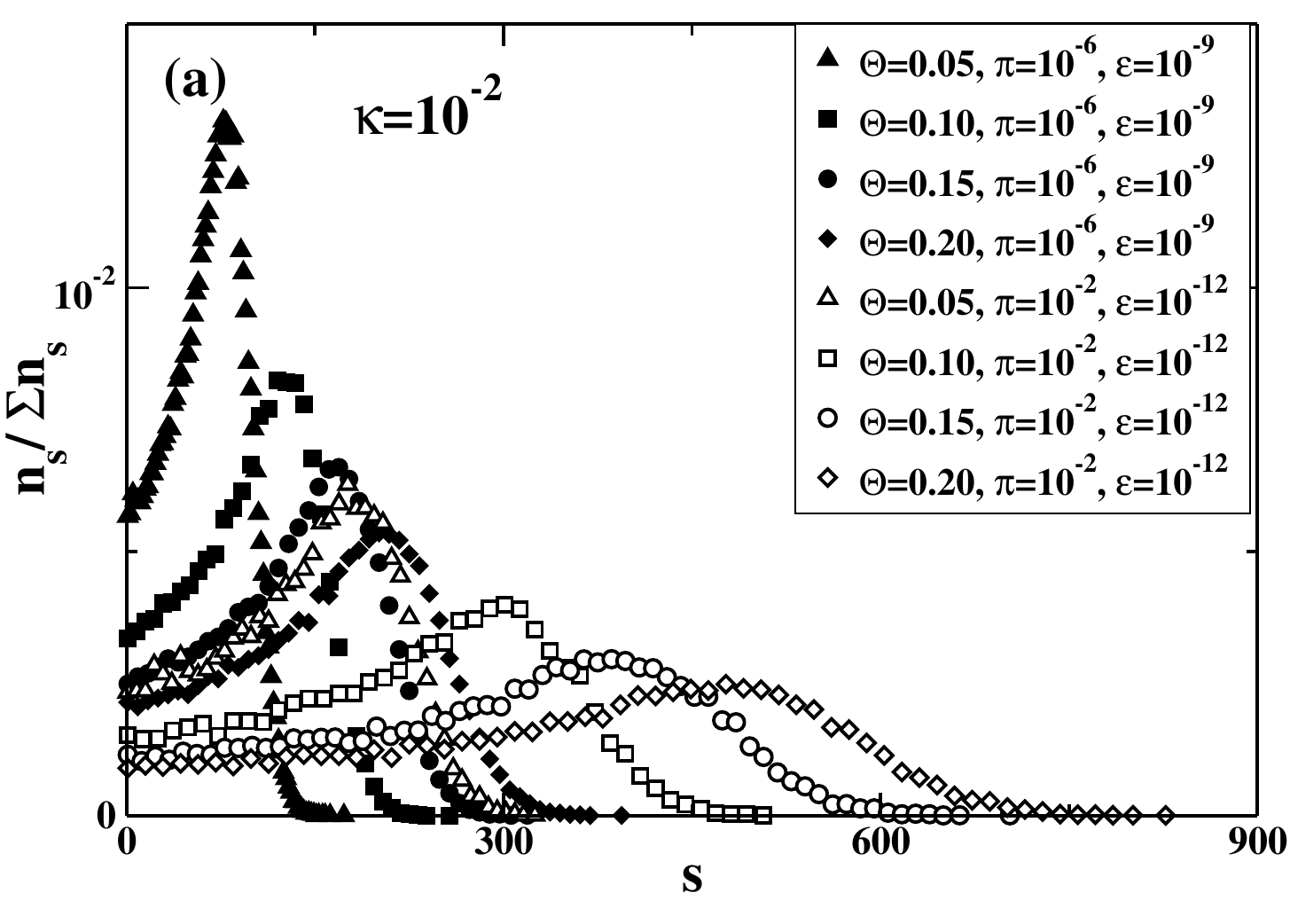}
\end{minipage}
\begin{minipage}[b]{0.5\linewidth}
\includegraphics[width=\linewidth]{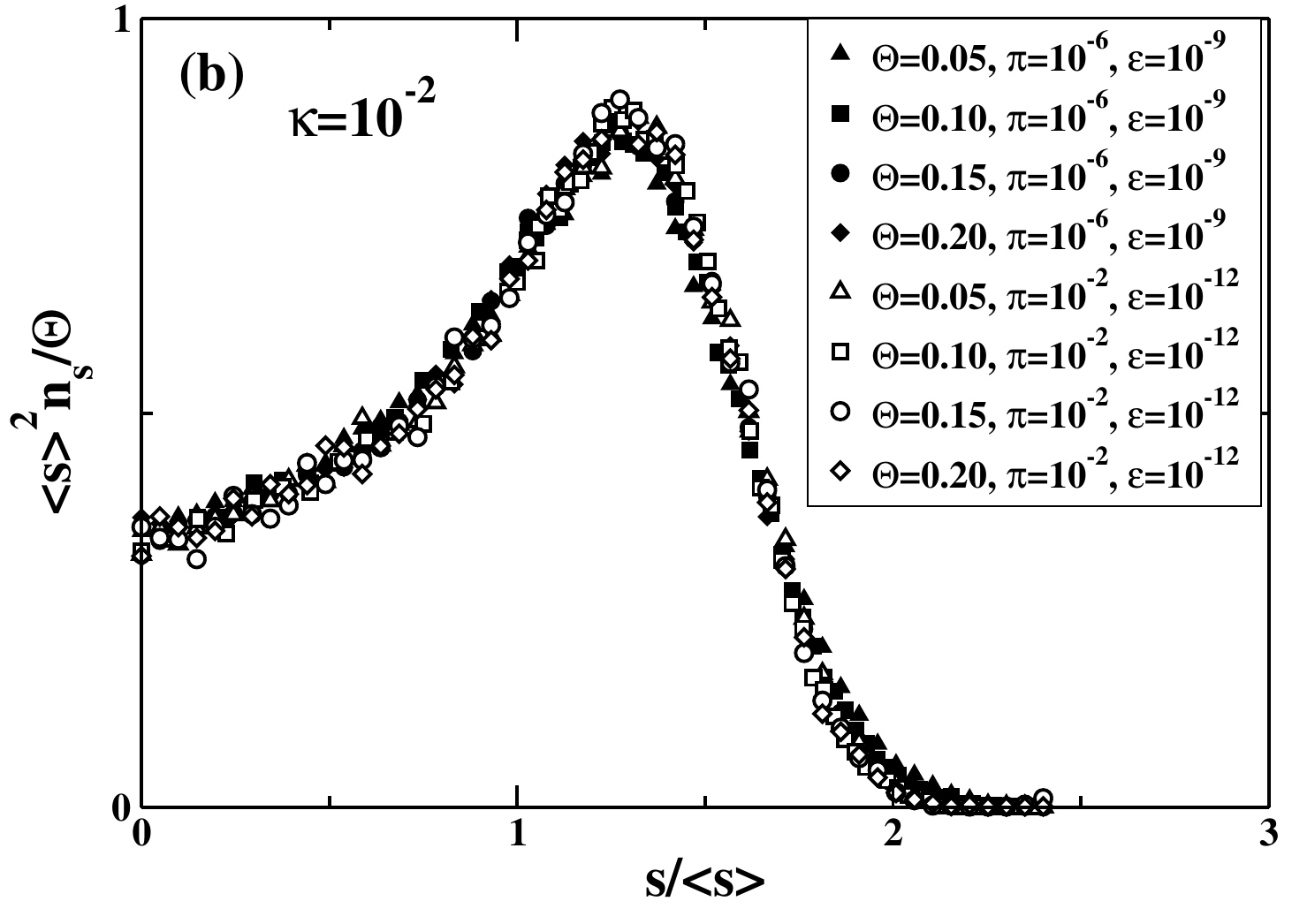}
\end{minipage}
~\vspace{1cm}\\
\begin{minipage}[b]{0.5\linewidth}
\includegraphics[width=\linewidth]{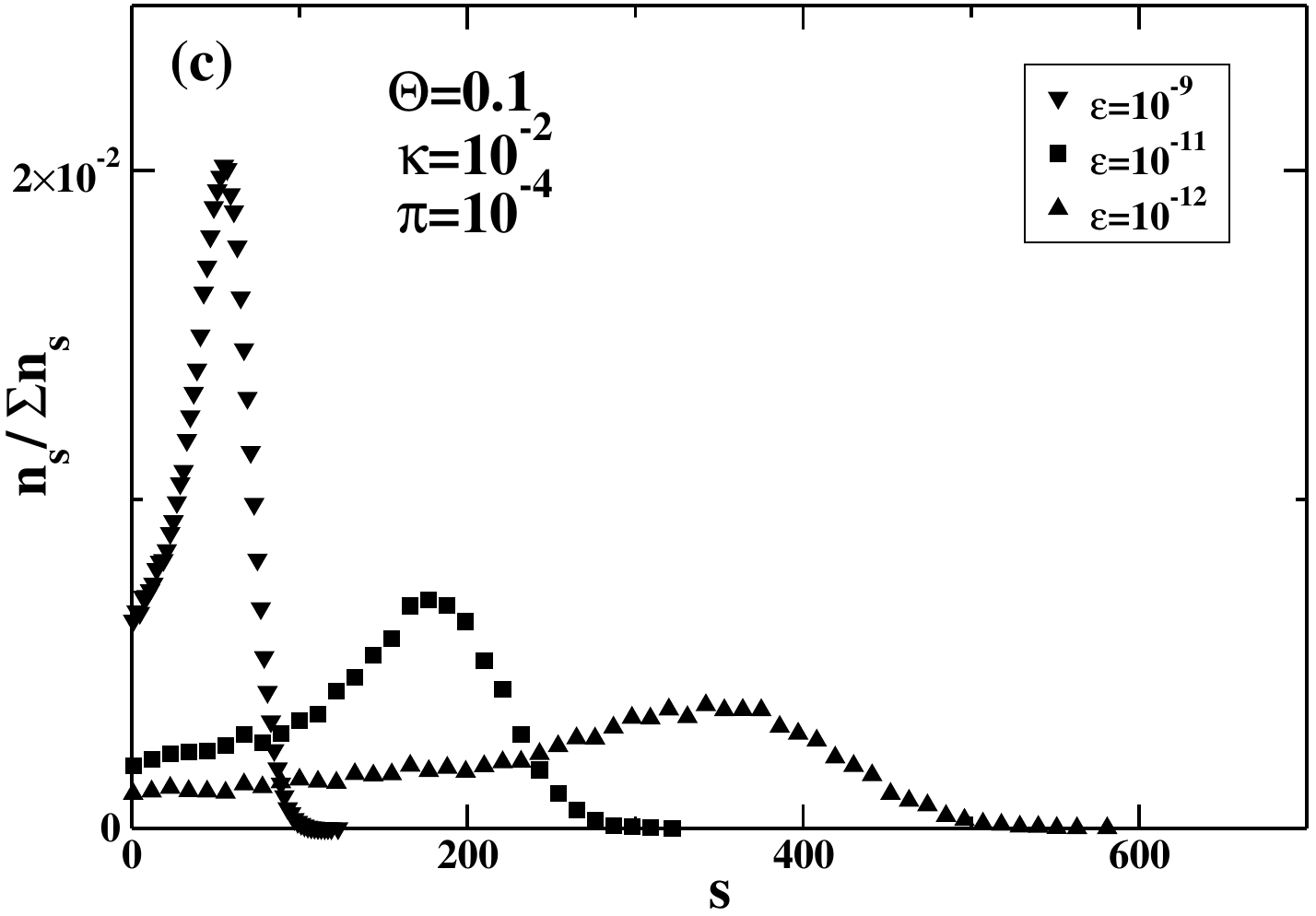}
\end{minipage}
\begin{minipage}[b]{0.5\linewidth}
\includegraphics[width=\linewidth]{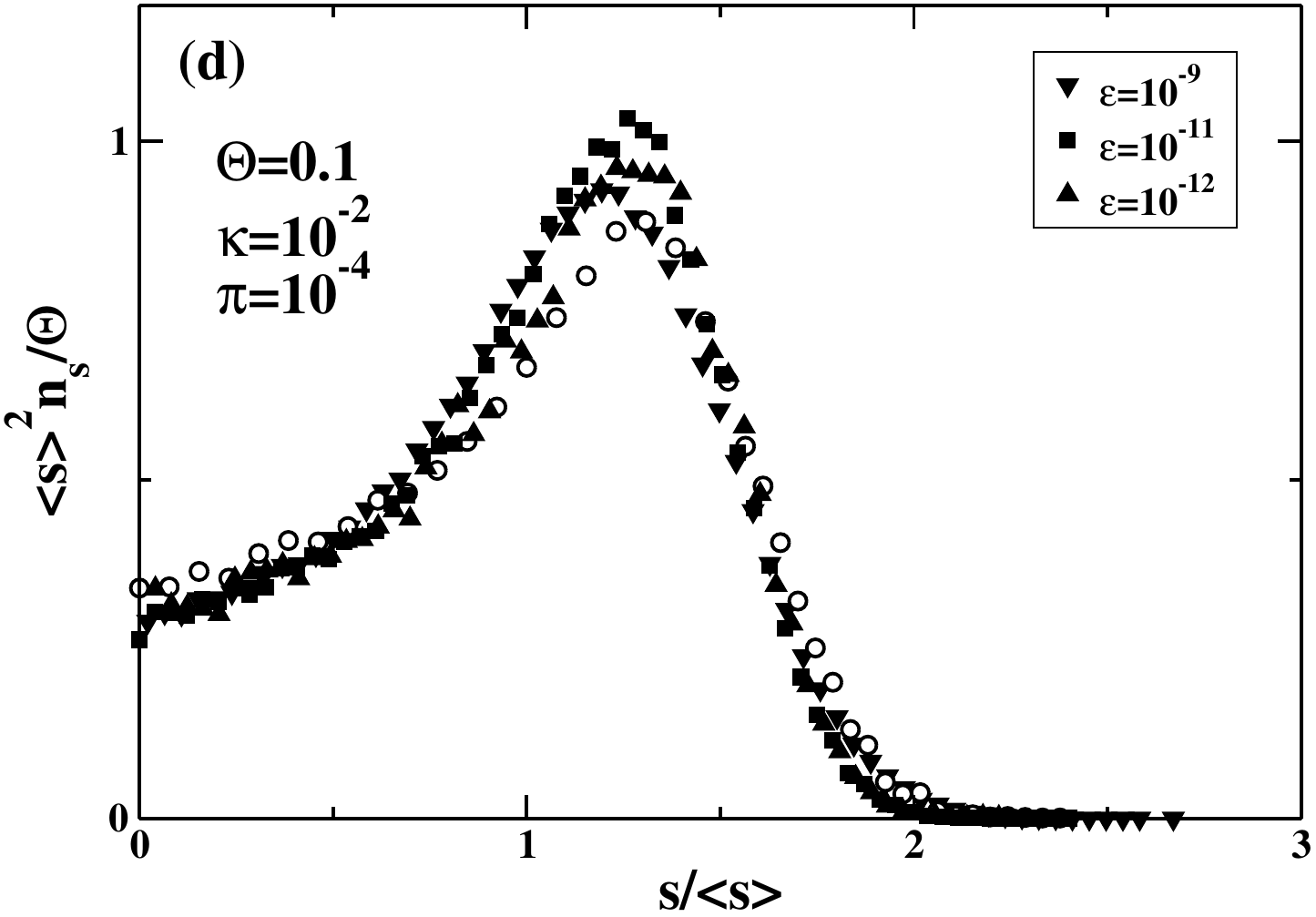}
\end{minipage}
\caption{Island size distributions. (a) Numerical results corresponding to the 
asymptotic regimes B-like ($\Theta_{\rm B}/\Theta > 0.99$,  solid symbols) and A-like 
($\Theta_{\rm B}/\Theta < 0.016$, open symbols). 
(b) Rescaled distributions for the same data in panel (a).
(c) Results of simulations for intermediate values of $\epsilon$; 
$\Theta_{\rm B}/\Theta=0.85$ (down triangles), $0.62$ (squares), and  
$0.50$ (up triangles). 
(d) Rescaled distributions corresponding to the functions in panel (c) 
(solid symbols), and  the scaling function from panel (b) (open symbols).}
\label{scale}
\end{figure}

As our model becomes the standard irreversible aggregation model when
$\epsilon\ll\epsilon_A$ or $\epsilon_B\ll\epsilon$, 
the island size distribution should satisfy the scaling hypothesis (\ref{scaling})
in these asymptotic regimes. 
To check this, we have performed Monte Carlo simulations, for the B-like case, with $\kappa=10^{-2}$, 
$\epsilon=10^{-9}$, and $\pi=10^{-6}$, which leads to $\Theta_{\rm B}/\Theta<0.016$. 
For the A-like case, we have chosen $\kappa=10^{-2}$, $\epsilon=10^{-12}$,  
and $\pi=10^{-2}$, which results in $\Theta_{\rm B}/\Theta>0.99$. 
 In figure \ref{scale}(a), we have plotted with solid symbols the numerical 
island size  distributions which correspond to the first group,
and with open symbols those which correspond to second;
at the coverages indicated in the figure key. The plots of $n_{\rm s} \left < s \right >^2/\Theta$, as a function of $s/\left < s \right >$, for the same data,
are shown in figure \ref{scale}(b). The good collapse of the data points
on a single curve is apparent, and gives support to the scaling law (\ref{scaling}) when only one diffusion constant is relevant to the dynamics.

For intermediate values of $\epsilon$, it is expected that the presence of a 
new rate, introduced with a second constant of diffusion, invalidates the 
scaling form (\ref{scaling}); in analogy to the case of detachment, when  $f$ is affected 
because of the rate related to this process~\cite{Vvedensky2000}.
In figure \ref{scale}(c) we shown three numerical island size distributions
for $\epsilon$ between $\epsilon_A$ and $\epsilon_B$, for which 
$\Theta_{\rm B}/\Theta=0.85, 0.62, 0.50$. The corresponding scaled 
functions are plotted in figure \ref{scale}(d); we have also included
the scaling function  from panel (b), for comparison.
Clear differences among all these functions are easily observed, which indicates
that the scaling behavior of the island size distribution
is indeed affected by the presence of  two species of
atoms moving on the substrate according to different diffusion constants.
}

%%%%%%%%%%%%%%%%%%%%%%%%%%%%%%%%%%%%%%%%%%%%%%%%%%%%%%%%%%%%%%%%%%%%%%%%%%%%%%%%

\section{Conclusions}
\label{Conclusions}
Despite the complexity and variety in reached morphologies of heteroepitaxial growths with intermixing, certain aspects of island growth appear to be common to many different systems. In the interest of archiving a complete and predictive model for the earliest stages of thin-film morphology that exhibit exchange between deposited and substrate atoms, it is clearly desirable to have an approach that is as free as possible from arbitrary parameters or assumptions. In this work, with an aim toward this ideal approach, we have presented a simple model to study the influence of intermixing and the different diffusion constants of the species moving on the surface, in island formation at low coverage. The model, only controlled by three parameters: the ratio between diffusion constants of the species, the non-dimensional incoming flux of particles and the non-dimensional intermixed probability of these particles with of substrate, can explain the behavior of density island and the variation of surface composition with 
time, for different values of these parameters. We found that the island dynamics are governed by the diffusion of the deposited atoms, at low temperature and by the diffusion of emerging particles from the substrate at high temperature regardless their diffusion constants.\par
We show that intermixing phenomenon is the predominant mechanism that can explain the island composition profile at low coverage. Then, other mechanisms interfere in this kind of thin film growths at higher coverage, such as the interactions between diffusing atoms.  Our model allows to study the effect of intermixing separated of the interactions between atoms tending to form 3D islands when the exposed surface of islands increases \cite{15Niehus99,16Niehus99,17Niehus99,18Niehus99,19Niehus99,20Niehus99}.\par

Mean-field evolution equations for island and monomer density have been written and resolved in simple situations, such as strong intermixing and high working temperatures and/or low deposition rates of atoms on the substrate. We found through these equations, a collapse of the island density for different values of the parameters of the model.\par 
\par

  Finally, we study the island-size distribution. The scaling 
behavior of this quantity is observed to be the same that for
the standard irreversible aggregation model, in the asymptotic regimes
where $\epsilon\ll\epsilon_A$ or $\epsilon_B\ll\epsilon$. In contrast, this scaling law fails at intermediate values of $\epsilon$, because of the  two species of atoms moving with different diffusion constants.

\ack This research was partially supported by a grant from the CONICET (PIP 0431), Universidad Nacional de Mar del Plata and ANPCyT (PICT),
Argentina. 

%%%%%%%%%%%%%%%%%%%%%%%%%%%%%%%%%%%%%%%%%%%%%%%%%%%%%%%%%%%%%%%%%%%%%%%%%%%%%

%%%%%%%%%%%%%%%%%%%%%%%%%%%%%%%%%%%%%%%%%%%%%%%%%%%%%%%%%%%%%%%%%%%%%%%%%%%%

\end{document}